\documentclass[aps,prb,twocolumn,showpacs,superscriptaddress,floatfix,10pt,nobibnotes]{revtex4-1}
\usepackage{bm,amsmath,amssymb,gensymb,amsfonts,stmaryrd,wasysym,graphicx,multirow,color,textcomp,braket}
\usepackage{url}
\usepackage{graphicx}
\usepackage{lipsum}
\usepackage[caption=false,singlelinecheck=false]{subfig}
\usepackage[colorlinks,linkcolor=blue,citecolor=cyan,filecolor=magenta,urlcolor=cyan]{hyperref}
\usepackage[super]{nth}

\newcommand{\nn}{\nonumber}
\definecolor{darkred}{RGB}{192, 0, 0}

\usepackage{multirow}    

\begin{document}

\title{
Spin Liquid Landscapes in the Kagome Lattice: A Variational Monte Carlo Study of the Chiral Heisenberg Model and Experimental Signatures
}

\author{Hee Seung Kim}
\email{shamoo0829@kaist.ac.kr}
\affiliation{Department of Physics, Korea Advanced Institute of Science and Technology, Daejeon, 34141, Republic of Korea}
\author{Hyeok-Jun Yang}
\email{hyang23@nd.edu}
\affiliation{Department of Physics, University of Notre Dame, South Bend, IN 46556}
\author{Karlo Penc}
\email{penc.karlo@wigner.hun-ren.hu}
\affiliation{HUN-REN Wigner Research Centre for Physics, P.O. Box 49, H-1525 Budapest, Hungary}
\author{SungBin Lee}
\email{sungbin@kaist.ac.kr}
\affiliation{Department of Physics, Korea Advanced Institute of Science and Technology, Daejeon, 34141, Republic of Korea}


\begin{abstract}
Chiral spin liquids, which break time-reversal symmetry, are of great interest due to their topological properties and fractionalized excitations (anyons). 
In this work, we investigate chiral spin liquids (CSL) on the kagome lattice arising from the competition between the third-nearest-neighbor Heisenberg interaction across hexagons ($J_d$) and a staggered scalar spin chirality term ($J_\chi$).
Using variational Monte Carlo methods, we map out the phase diagram and identify various gapped and gapless CSL phases, each characterized by a distinct flux pattern. 
Notably, the interplay between $J_d$ and $J_\chi$ induces a tricritical point, which we analyze using Landau-Ginzburg theory. 
Additionally, we identify potential signatures of these CSLs—including distinctive spin-spin correlations, anomalies in the static spin structure factor, longitudinal thermal conductivity, and magnetoelectric effects—which offer practical guidance for their future experimental detection.
\end{abstract}

\maketitle

\textit{Introduction--} Quantum spin liquids (QSL) are receiving significant attention in both theoretical and experimental research due to their exotic properties, such as long-range entanglement in the absence of any magnetic ordering \cite{Balents2010, Savary_2017, doi:10.1146/annurev-conmatphys-031218-013401, doi:10.1126/science.1163196, RevModPhys.89.025003}. 
Especially in the antiferromagnetic Heisenberg model on the kagome lattice, the precise nature of the QSL ground state is still under debate. Notable ground-state candidates include 
a $\mathbb{Z}_{2}$ gauge structure with a gapped spectrum~\cite{PhysRevB.83.224413, doi:10.1126/science.1201080,PhysRevLett.109.067201} and
a $U(1)$ gauge structure with a gapless Dirac dispersion~\cite{PhysRevLett.98.117205, PhysRevB.84.020407, PhysRevB.87.060405, PhysRevB.89.020407}.
Among the various QSLs on the kagome lattice, the chiral spin liquid (CSL), which breaks the time-reversal symmetry $\mathcal{T}$ (and parity symmetry depending on the lattice structure)  has gained particular interest due to its topologically protected edge modes, quantized thermal Hall conductance, and anyonic statistics \cite{Rong-Yang_chiralKagome_2024, PhysRevB.92.060407, PhysRevLett.99.097202, PhysRevLett.99.247203, PhysRevB.80.104406, PhysRevLett.110.067208, PhysRevLett.112.137202, PhysRevLett.115.267209, PhysRevB.95.035141, PhysRevX.10.021042, PhysRevLett.116.137202, Bauer2014, PhysRevB.99.035155, 10.21468/SciPostPhys.13.3.050, bauer2014gapped, 10.21468/SciPostPhys.4.1.004, PhysRevB.92.125122}. Recently, density matrix renormalization group and variational Monte Carlo (VMC) studies on the kagome lattice found evidence for a CSL by considering an extended Heisenberg model up to third neighbor interaction on the kagome lattice~\cite{PhysRevB.91.075112, Gong2014, PhysRevLett.112.137202}. According to Ref.~\onlinecite{Rong-Yang_chiralKagome_2024}, this CSL also includes the case with only nearest-neighbor Heisenberg exchanges.

There are two well-known CSLs on the kagome lattice: the Kalmeyer-Laughlin type, which breaks reflection symmetry \(\sigma\) but preserves \(\sigma\mathcal{T}\), and the staggered flux type, which breaks the six-fold rotation symmetry \(C_{6}\) but preserves \(C_{6}\mathcal{T}\)~\cite{PhysRevLett.59.2095, PhysRevB.93.094437}. 
While these two types of CSLs are well-known, the kagome lattice allows for numerous other CSL classes, each characterized by different projective symmetry groups~\cite{PhysRevB.92.060407}.
However, only a few microscopic spin Hamiltonians are known to realize these CSL classes as ground states on the kagome lattice.

In this paper, we revisit the spin-1/2 $J_{1}$–$J_{d}$–$J_{\chi}$ microscopic model on the kagome lattice~\cite{10.21468/SciPostPhys.13.3.050}, using variational Monte Carlo (VMC) methods to investigate the realization of different classes of chiral spin liquids (CSLs).
The Hamiltonian of the model is given by
\begin{align}
\label{eq:J1JdJcHamiltonian}
    \hat{\mathcal{H}} 
    &= J_{1}\sum_{\braket{i,j}}\hat{\bm{S}}_{i}\cdot\hat{\bm{S}}_{j} + J_{d}\sum_{\braket{\braket{\braket{i,j}}}\in\hexagon}\hat{\bm{S}}_{i}\cdot\hat{\bm{S}}_{j} \\
    &+ J_{\chi} \Big[\sum_{ijk\in\vartriangle} \hat{\bm{S}}_{i}\cdot(\hat{\bm{S}}_{j}\times\hat{\bm{S}}_{k}) 
    -  \sum_{ijk\in\triangledown} \hat{\bm{S}}_{i}\cdot(\hat{\bm{S}}_{j}\times\hat{\bm{S}}_{k}) \Big] \nonumber,
\end{align}
where $\hat{\bm{S}}_{i}$ is a spin-1/2 operator at site $i$, $J_{1} =1 $ is the nearest-neighbor and $J_{d} > 0$ the third-neighbor antiferromagnetic Heisenberg exchange across the hexagons. 
To incorporate three-spin correlations, we add the scalar spin-chirality term, $J_{\chi}$. The sum runs over every elementary triangular plaquette of the lattice, with the sites $i$, $j$, and $k$  listed in clockwise order. Upward-pointing ($\vartriangle$) and downward-pointing ($\triangledown$) triangles enter with opposite signs, thereby generating a staggered pattern of chirality across the lattice
\footnote{At the microscopic level, the staggered spin chiral interaction arises from anisotropic exchanges induced by spin-orbit coupling and from the influence of external magnetic fields under conditions of reduced symmetry. The forthcoming paper will elaborate on these specific mechanisms and details.
}.
The $J_{\chi}$ term breaks both $\mathcal{T}$ and $C_{6}$ individually, while their combination $C_{6}\mathcal{T}$ remains a symmetry.

\begin{figure*}
	\centering
	\includegraphics[width=\textwidth]{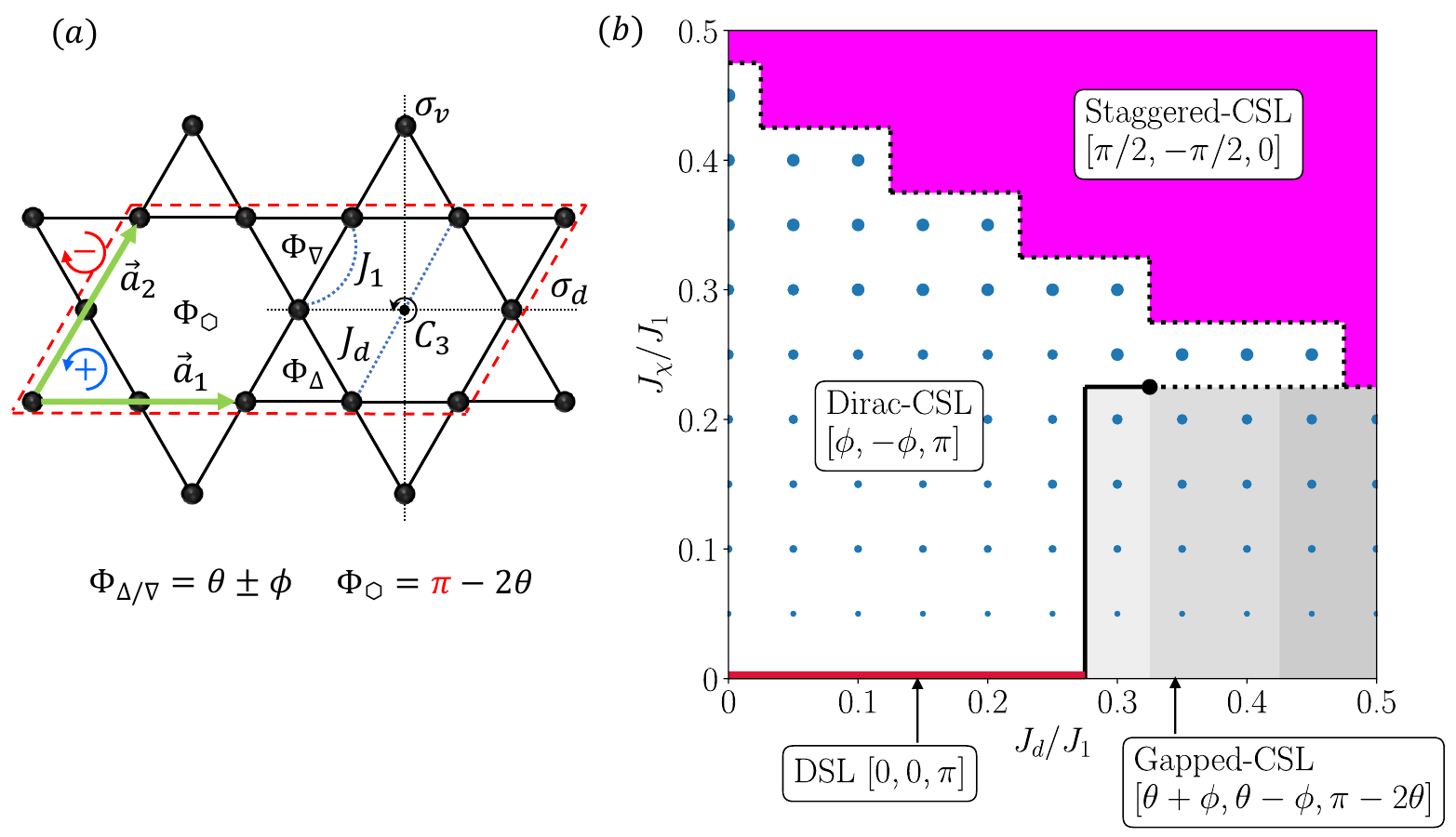}
	\caption{ 
(a) The flux pattern of Eq.~\eqref{eq:J1JdJcMeanHamiltonian} on the kagome lattice. The $U(1)$ gauge fluxes are $\Phi_{\vartriangle} = \theta + \phi$ and $\Phi_{\triangledown} = \theta - \phi$ through the triangles and $\Phi_{\hexagon} = -2\theta$ through the hexagons.
The red dotted line indicates the enlarged unit cell of the type-1 flux pattern. The fermionic spinon hopping amplitude is $|t_{ij}| = 1$ for all nearest-neighbor pairs.
(b) The variational phase diagram of the $J_{1}-J_{d}-J_{\chi}$ model for a $N=3 \times 12 \times 12$ site cluster, constructed by comparing energies calculated with VMC for competing Gutzwiller-projected states with the flux parameters \( \theta \) and  \( \phi \) optimized at each point in parameter space. Within this framework, four stable phases appear: the Dirac spin liquid (DSL), the Dirac chiral spin liquid (Dirac CSL), the gapped chiral spin liquid (gapped CSL), and the staggered chiral spin liquid (staggered CSL). The radius of the blue circles encodes the magnitude of the staggered flux  \( \phi \), while the intensity of the gray shading is proportional to \( \theta \), where \( \Phi_{\hexagon} = \pi - 2\theta \). In the magenta region \( \phi = \pi/2 \), \( \theta = 0 \), and \( \Phi_{\hexagon} = 0 \). The flux patterns and point group symmetries of each phase are detailed in Table~\ref{tab:tab}. Solid lines represent second-order phase transition boundaries, while dotted lines indicate first-order transitions. 
Turning on $J_{\chi}$ immediately leads to time-reversal symmetry breaking and drives a phase transition from the DSL to the Dirac-CSL.}
\label{fig:KagomeLattice}
\end{figure*}

\textit{Variational Monte Carlo---} To construct possible QSL states of Hamiltonian \eqref{eq:J1JdJcHamiltonian}, we start with a fermionic mean-field Hamiltonian,
\begin{align}
    \label{eq:J1JdJcMeanHamiltonian}
    \hat{\mathcal{H}}_{\text{MF}} = \sum_{\braket{ij},\alpha}\left(t_{ij}\hat{f}^{\dagger}_{i\alpha}\hat{f}_{j\alpha} + h.c.\right). 
\end{align}
Here $\hat{f}_{i\alpha}$ are fermionic spinon operator that represents the spin operator as 
$\hat{\bm{S}}_{i} = \frac{1}{2}\sum_{\alpha,\beta = \uparrow, \downarrow}\hat{f}_{i\alpha}^{\dagger}\bm{\sigma}_{\alpha\beta}\hat{f}_{i\beta}$,
where 
$\bm{\sigma} = (\sigma^{x}, \sigma^{y}, \sigma^{z})$ 
is a vector of Pauli matrices, and 
$\hat{n}_{i} = \sum_{\alpha} \hat{f}_{i\alpha}^{\dagger}\hat{f}_{i\alpha}$ counts the number of spinons at site $i$.
The ground state of the mean-field Hamiltonian is a Fermi sea of spinons, denoted as $\ket{\Psi_{\text{MF}}}$. 
We consider a half-filled system with the number of spinons equal to the number of sites, $\bra{\Psi_{\text{MF}}}\sum_{i=1}^{N}\hat{n}_{i}\ket{\Psi_{\text{MF}}} = N$.
To eliminate the spinon-number fluctuations, we use the variational wave function
$\ket{\Psi_{G}} \equiv P_{G}\ket{\Psi_{\text{MF}}}$, 
where the Gutzwiller projection $P_{G} = \prod_{i}\hat{n}_{i}(2-\hat{n}_{i})$
ensures that the half-filling condition is satisfied at each site.
 We focus on different compact $U(1)$ gauge flux patterns to characterize the QSL phases.
  Specifically, we consider uniform nearest-neighbor hopping amplitudes with $|t_{ij}| = 1$ and minimize the variational energy, $\braket{\Psi_{G}|\hat{\mathcal{H}}|\Psi_{G}}$ with respect to the $U(1)$ gauge fluxes through the lattice.

Each unit cell comprises three plaquettes: the up- and down-pointing triangles and the hexagon. We introduce two parameters to characterize the chiral QSL states: a uniform flux $\theta$ and a staggered flux $\phi$ through the triangles. Then, the total flux through the up triangles is $\Phi_{\vartriangle} = \theta + \phi$ and the down triangles is $\Phi_{\triangledown} = \theta - \phi$.
The flux of the hexagonal plaquettes, $\Phi_{\hexagon}$, can take one of two forms depending on the type of flux pattern: for type-1, $\Phi_{\hexagon} = \pi - 2\theta$; for type-2, $\Phi_{\hexagon} = -2\theta$ (see Fig.~\ref{fig:KagomeLattice}(a)). For type-1, the total flux threading the elementary unit cell is $\pi$, requiring a doubled unit cell size in our calculations.
By specifying the gauge fluxes on these three plaquettes, we can uniquely distinguish different QSL states, which we denote as SL[$\Phi_{\vartriangle}, \Phi_{\triangledown}, \Phi_{\hexagon}$].

We performed VMC calculations on a finite lattice with $N = 3 \times 12 \times 12$ sites and obtained the phase diagram shown in Fig.~\ref{fig:KagomeLattice}(b). We repeated the calculations at selected parameter points for $N = 3 \times n \times n$ sites, with $n = 4, 8, 12, 16$. In all cases, the CSL consistently emerged as the lowest-energy state within the set of variational states considered, indicating that the phase diagram is robust against finite-size effects.

The finite-size mean-field spectrum depends on gauge fluxes  $(\theta, \phi)$, which can induce Fermi-level degeneracies and render the half-filled ground state  $\ket{\Psi_{\text{MF}}}$ ambiguous. We addressed this using twisted boundary conditions, which shift the momentum grid but break $C_6$ symmetry, causing the Gutzwiller-projected wave function to break $C_6$ likewise and bias observables such as spin correlations. We recovered symmetry by superposing projected states related by point-group operations (see Appendix A).
 
\begin{figure*}
	\centering
	\includegraphics[width = \textwidth]{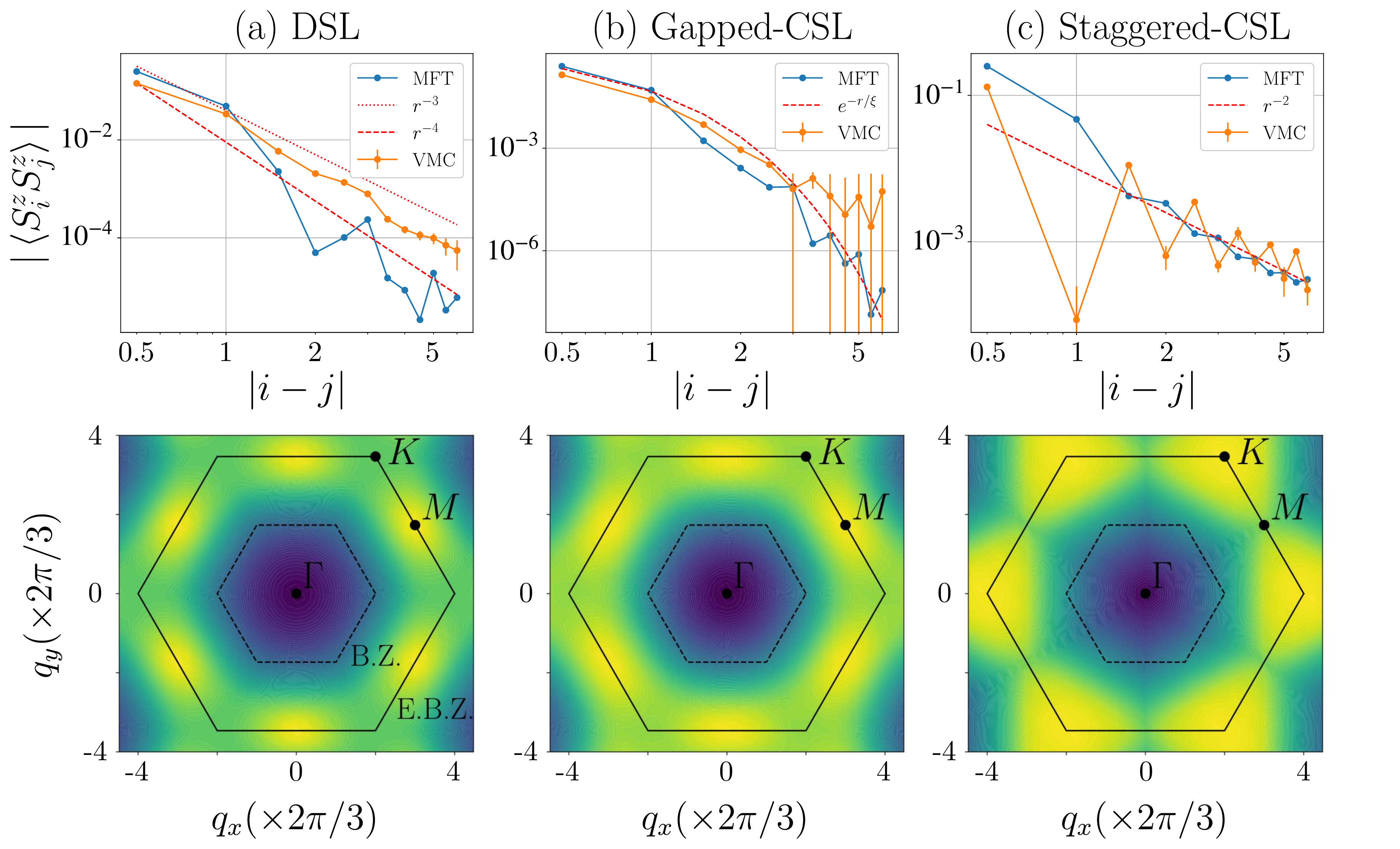}
	\caption{(Top) Spin-spin correlation of (a) Dirac spin liquid ($\Phi_\vartriangle = 0$, $\Phi_\triangledown=0$, $\Phi_{\hexagon}=\pi$), (b) gapped chiral spin liquid ($\Phi_\vartriangle = \pi/8$, $\Phi_\triangledown = \pi/8$, $\Phi_{\hexagon}=3\pi/4$), and (c) staggered chiral spin liquid ($\Phi_\vartriangle=\pi/2$, $\Phi_\triangledown=-\pi/2$, $\Phi_{\hexagon}=0$), where ($\Phi_\vartriangle$, $\Phi_\triangledown$, $\Phi_{\hexagon}$) denote the fluxes through the up and down triangles and hexagons of the kagome lattice, respectively. We plot variational Monte Carlo (VMC) and mean-field results for each case. (Bottom) Corresponding static spin structure factor. The dotted and solid lines represent the 1st and the extended Brillouin zone, respectively. The structure factor is normalized over the extended Brillouin zone, $\sum_{\bm{q}\in \text{EBZ}}S(\bm{q}) = 1$.}
	\label{fig:staticspinstructurefactor}
\end{figure*}

\begin{table*}[bt]
\caption{
Flux patterns $\mathrm{SL}[\Phi_{\vartriangle},\Phi_{\triangledown},\Phi_{\hexagon}]$, real-space decay of the $\lvert S_i^{z} S_j^{z}\rvert$ spin--spin correlations and low-temperature behavior of the thermal conductivity $\kappa_{xx}$ , magnetic point groups, and magnetoelectric invariants for the different flux states on the kagome lattice and Kapellasite. 
Table entries of the form ``\checkmark/-'' indicate whether the listed term is an invariant for the kagome lattice (first symbol) and for Kapellasite (second symbol), with \checkmark (-) denoting allowed (forbidden) invariants.}
\label{tab:tab}
\begin{ruledtabular}
\begin{tabular}{cccccc}
&
DSL &
Dirac CSL&
staggered CSL&
gapped CSL &
gapped CSL \\
\hline
SL[$\Phi_{\vartriangle}$,$\Phi_{\triangledown}$,$\Phi_{\hexagon}$] & 
SL[$0$,$0$,$\pi$] & SL[$\phi$,$-\phi$,$\pi$] & SL[$\pi/2$,$-\pi/2$,$0$] & SL[$\theta$,$\theta$,$\pi-2\theta$]  & SL[$\theta+\phi$,$\pi-2\theta$] 
\medskip\\
$\left|S_{i}^{z}S_{j}^{z}\right|$ & \multicolumn{2}{c}{$|i-j|^{-\alpha}~(3\leq\alpha\leq 4)$} & $|i-j|^{-2}$ & Exp. decay &Exp. decay \\
$\kappa_{xx}$ & Power law  &  \multicolumn{2}{c}{Power law}  & Exp. decay & Exp. decay 
\medskip\\
Symmetry group -- kagome &
$\{1,\mathcal{T}\}\otimes \mathsf{D_{6h}}$ &
\multicolumn{2}{c}{$\{1,\mathcal{T}\sigma_v\}\otimes \mathsf{D_{3h}}$}&
$\{1,\mathcal{T}\sigma_v\}\otimes \mathsf{C_{6h}}$ &
$\{1,\mathcal{T}\sigma_v\}\otimes \mathsf{C_{3h}}$ 
\\Symmetry group -- Kapellasite&
$\{1,\mathcal{T}\}\otimes \mathsf{D_{3d}}$ &
\multicolumn{2}{c}{$\{1,\mathcal{T}\sigma_v\}\otimes \mathsf{D_{3}}$}&
$\{1,\mathcal{T}\sigma_v\}\otimes \mathsf{S_{6}}$ &
$\{1,\mathcal{T}\sigma_v\}\otimes \mathsf{C_{3}}$ 
\medskip\\
$P_y^3-3 P_x^2 P_y$, $2 P_x M_x M_y+P_y (M_x^2-M_y^2)$ & -/- & \multicolumn{2}{c}{-/-}  & -/- &\checkmark/\checkmark \\ 
$P_z$, $M_z (P_x M_x+P_y M_y)$ & -/- & \multicolumn{2}{c}{-/-}  & -/- & -/\checkmark 
 \medskip\\
$M_z$, $(P_x M_x+P_y M_y) P_z $ & -/- &\multicolumn{2}{c}{-/-} &\checkmark/\checkmark &\checkmark/\checkmark \\
$2 P_x P_y M_x + (P_x^2 - P_y^2) M_y$, $ M_y^3-3 M_x^2 M_y$ &  -/-  &  \multicolumn{2}{c}{-/-}  &  -/\checkmark  &  -/\checkmark  
 \medskip\\
$P_z \left[2 P_x P_y M_x + (P_x^2-P_y^2) M_y\right]$& -/- & \multicolumn{2}{c}{\checkmark/\checkmark} & -/- &\checkmark/\checkmark\\
$M_z \left[2 M_x M_y P_x + (M_x^2-M_y^2) P_y\right]$& -/- & \multicolumn{2}{c}{\checkmark/\checkmark} & -/- &\checkmark/\checkmark\\
$ \left(M_y^3-3 M_x^2 M_y\right)  P_z$, $ \left(P_y^3-3 P_x^2 P_y\right)  M_z$& -/- & \multicolumn{2}{c}{\checkmark/\checkmark} & -/- &\checkmark/\checkmark\\
$P_x M_x+P_y M_y$, $P_z M_z$ &  -/-  &  \multicolumn{2}{c}{-/\checkmark}  &  -/-  &  -/\checkmark  
\medskip\\
$P_x^2+P_y^2$, $P_z^2$, $M_x^2+M_y^2$, $M_z^2$  &\checkmark/\checkmark&\multicolumn{2}{c}{\checkmark/\checkmark} 
&\checkmark/\checkmark&\checkmark/\checkmark\\
\end{tabular}
\end{ruledtabular}
\end{table*}%

\textit{Phase diagram---} 
In the absence of chiral term ($J_{\chi} = 0$), the time-reversal invariant SL[$0$, $0$, $\pi$] state minimizes the energy for small $J_{d} < J_{d}^{c} \approx 0.25 \sim 0.3$.
The spinon band structure features two Dirac cones within the reduced Brillouin zone, hence the name ``Dirac spin liquid'' (DSL) \cite{PhysRevLett.98.117205, PhysRevB.87.060405, PhysRevB.89.020407, PhysRevB.77.224413}. These cones are protected by the mirror symmetry $\sigma_d$~\cite{PhysRevB.93.094437}. 
Although the $\pi$ flux doubles the unit cell, the projective symmetry group restores the full translational and $D_{6h}$ point group symmetries of the kagome lattice.

As $J_{d}$ increases above the critical value $J_{d}^{c}$, the system stabilizes a state that spontaneously breaks the time-reversal symmetry, the CSL with the flux pattern labeled by SL[$\theta$, $\theta$, $\pi - 2\theta$]. Since this flux pattern also breaks the $\sigma_d$ symmetry that protects the Dirac cones, a gap opens inside the bulk, giving the bands finite Chern numbers. This state, known as the gapped chiral spin liquid (gapped CSL), is analogous to a fractional quantum Hall state with filling factor $\nu = 1/2$~\cite{Gong2014, PhysRevB.91.041124}. 

A finite $J_{\chi}$ in Eq.~\eqref{eq:J1JdJcHamiltonian} explicitly breaks $\mathcal{T}$ symmetry in a staggered manner, favoring opposite gauge fluxes on the up and down triangles and resulting in a nonzero staggered flux $\phi \neq 0$. 
Consequently, this leads to a phase transition from the DSL into a distinct CSL characterized by the flux pattern SL[$\phi$, $-\phi$, $\pi$].
Although this flux pattern breaks $\mathcal{T}$ symmetry, it preserves horizontal reflection symmetry, keeping the Dirac cones gapless. This state is called the Dirac chiral spin liquid (Dirac CSL). 
As $J_{\chi}$ increases, the staggered chiral flux $\phi$ in the Dirac spin liquid (DSL) grows continuously until the system undergoes a first-order transition into the staggered chiral spin liquid phase, discussed below.
According to Table IX of Ref.~\onlinecite{PhysRevB.93.094437}, Dirac CSL corresponds to No.~12 among staggered flux $U(1)$ CSL phases, marking its first realization of non-Fermi liquids with Dirac Fermi points of fractionalized excitations.

Similarly, starting from a gapped CSL, a staggered flux $\phi$ appears as $J_{\chi}$ increases, changing the flux pattern from SL[$\theta$, $\theta$, $\pi - 2\theta$] to SL[$\theta + \phi$, $\theta - \phi$, $\pi - 2\theta$]. Although a finite $\phi$ reduces the rotational symmetry from $C_{6}$ to $C_{3}$, it preserves the spinon gap structure and the associated fractional quantum Hall state. Therefore, we consider this the same gapped CSL phase as the $\phi = 0$ case for $J_{\chi} = 0$. However, the electromagnetic response will differ due to the reduced symmetry, as we will discuss in the experimental probes section.

For small values of $J_{\chi}$, the phase transition between the Dirac spin liquid (DSL) and the gapped chiral spin liquid (CSL) is continuous. However, as $J_{\chi}$ increases further, this transition becomes discontinuous (first-order). The tricritical point where the nature of the transition changes from continuous to discontinuous is indicated by the black circle in Fig.\ref{fig:KagomeLattice} (b). We provide a detailed Landau-Ginzburg analysis of these phase transitions in terms of the parameters $\theta$ and $\phi$ in Appendix \ref{sec:tricritical}.

For sufficiently large $J_{\chi}$, the fluxes on the triangles reach their maximum values, $\Phi_{\vartriangle} = -\Phi_{\triangledown} = \pi/2$, while the flux on the hexagon stabilizes at $\Phi_{\hexagon} = 0$ (type-2) rather than $\Phi_{\hexagon} = \pi$ (type-1). We refer to this SL[$\pi/2$, $-\pi/2$, $0$] state as the staggered chiral spin liquid (staggered CSL). Its mean-field spectrum is gapless and exhibits line degeneracies at the Fermi level along the vertical reflection symmetry lines, protected by the anti-commutation relation $\{\hat{\mathcal{H}}_{\text{MF}}, \sigma_{v}\} = 0$~\cite{PhysRevB.99.035155}, as discussed in Appendix \ref{sec:fermilines}. In Ref.~\onlinecite{10.21468/SciPostPhys.13.3.050}, the authors suggested a purely imaginary nearest-neighbor hopping mean-field amplitude in the fermionic spinon basis for the large $J_{\chi}$ limit as a prominent example of non-Fermi liquids. This mean-field approach generates $\pm\pi/2$ fluxes through the up and down triangles.
Our calculations confirm that this staggered flux pattern with $\Phi_{\vartriangle} = -\Phi_{\triangledown} = \pi/2$ yields the minimum variational energy, even if we treat the flux as a variational parameter for large $J_{\chi}$. The staggered CSL corresponds to $U(1)$ CSL phase No.~11 in Table IX of Ref.~\onlinecite{PhysRevB.93.094437}. 

\textit{Experimental probes---} We begin by presenting the absolute value of the spin-spin correlation function, $\left|\braket{S_{i}^{z}S_{j}^{z}}\right|$, for (a) DSL, (b) gapped CSL, and (c) staggered CSL calculated by VMC simulations in Fig.~\ref{fig:staticspinstructurefactor}. 
Being gapless, both the DSL and the staggered CSL exhibit algebraic decay of spin correlations, $\left|\braket{S_{i}^{z}S_{j}^{z}}\right| \propto |i-j|^{-\alpha}$, but with different exponent: $3 \leq \alpha \leq 4$ in the DSL~\cite{PhysRevB.77.224413} and $\alpha \approx 2$ in the staggered CSL. We also compare the real-space decay of the spin correlations with predictions from mean-field theory in Fig.~\ref{fig:staticspinstructurefactor}(a) and (c). In contrast, the spin correlations decay exponentially in the gapped CSL, as shown in Fig.~ \ref{fig:staticspinstructurefactor}(b).

The bottom row of  Fig.~\ref{fig:staticspinstructurefactor} displays the static spin structure factor $S^{zz}(\bm{q})$, 
\begin{align}
	\label{eq:ssf}
	S^{zz}(\bm{q}) = \sum_{i,j}e^{i\bm{q}\cdot(\bm{r}_{i} - \bm{r}_{j})}\braket{S_{i}^{z}S_{j}^{z}}.
\end{align}
The spin structure factor of the DSL exhibits peaks at the $M$ points of the extended Brillouin zone~\cite{PhysRevB.77.224413}, consistent with classical calculations of the ``$q=0$ phase''~\cite{PhysRevB.91.104418}. In the gapped CSL, the peaks remain at the $M$ points but become broadened due to the spinon gap opening [Fig.\ref{fig:staticspinstructurefactor}(b)]. In contrast, the staggered CSL shows a fundamentally different behavior, with the structure factor peaking at the $K$ points [Fig.\ref{fig:staticspinstructurefactor}(c)].
Notably, in the presence of the staggered chirality term, classical spins order in a so-called cuboc-1 state, which similarly features peaks near the $K$ points~\cite{10.21468/SciPostPhys.13.3.050}.

Next, we investigate thermal conductivity as a sensitive probe of the excitations and their dispersions, since these excitations facilitate heat transport. 
At the mean-field level, ignoring gauge fluctuations, spinons are the primary degrees of freedom. We thus focus on the low-temperature behavior of the longitudinal thermal conductivity tensor
$\kappa_{xx}$~\cite{MOUSAVI20163823, PhysRevB.67.115131, PhysRevB.69.165105}. In the gapless Dirac CSL and staggered CSL phases, $\kappa_{xx}$ exhibits a power-law temperature dependence. In contrast, the longitudinal thermal conductivity of the gapped CSL decays exponentially with decreasing temperature, with the decay rate determined by the spinon gap.

Finally, we consider the response to an electric field.
For finite values of $\theta$ and $\phi$, the chiral spin liquid breaks the grey magnetic point group $\{1,\mathcal{T}\}\otimes \mathsf{D_{6h}}$ 
of the kagome lattice, i.e., the direct product of the crystallographic point group $\mathsf{D_{6h}}$ with the group generated by time-reversal symmetry $\mathcal{T}$, as illustrated in Fig.~\ref{fig:magnetic_point_groups}. The reduced symmetry permits a magnetoelectric coupling between the magnetization $\mathbf{M}=(M^x,M^y,M^z)$, with $\mathbf{M}=\sum_i \mathbf{S}_i$, and the electric polarization $\mathbf{P}=(P^x,P^y,P^z)$. Table~\ref{tab:tab} lists the symmetry-allowed invariants. 
For example, at finite $\phi$, the $ \left(M_y^3-3 M_x^2 M_y\right)  P_z$ is allowed, implying that rotating an external magnetic field within the kagome plane by an angle $\eta$ induces an out-of-plane electric polarization $P_z \propto \cos 3\eta$. 
This behavior is analogous to the magnetoelectric response observed in Ba$_2$CoGe$_2$O$_7$~\cite{Murakawa_PhysRevLett.105.137202}.
More generally, the set of symmetry-allowed invariants provides a direct diagnostic of the broken symmetries and, consequently, of the underlying flux structure of the spin liquid, as discussed in Appendix~\ref{sec:magnetoelectric}. For comparison, we also evaluate the magnetoelectric couplings for Kapellasite, whose point-group symmetry $\mathsf{D_{3d}}$ is lower than that of the ideal kagome lattice $\mathsf{D_{6h}}$. 
In this case, the invariants $P_{x}M_{x}+P_{y}M_{y}$ and $P_{z}M_{z}$ are allowed once $J_{\chi}\neq 0$, implying that an external magnetic (electric) field can linearly induce electric (magnetic) polarization in the gapless CSL\cite{PhysRevLett.119.077206}.

We also explored the microscopic origin of the magnetoelectric coupling.
A finite scalar chirality emerges in a third-order hopping process in a Mott insulator when a finite flux threads through a triangle and couples via the Peierls phase to the charges \cite{Motrunich_PhysRevB.73.155115}. 
Similarly, asymmetries in the bond expectation values $\langle \mathbf{S}_i \cdot \mathbf{S}_j \rangle$ induce charge imbalance and electric polarization within the kagome plane \cite{Bulaevskii_PhysRevB.78.024402}. 
To investigate whether these mechanisms capture the phenomenological responses, we performed exact diagonalization on a small 12-site cluster. 
Upon rotating an in-plane electric field, we found a varying total scalar chirality but no induced magnetization.
Thus, to obtain finite $P_z$ and magnetization $\mathbf{M}$, it becomes necessary to include spin-orbit coupling effects such as the spin-current mechanism~\cite{Katsura_PhysRevLett.95.057205, Sergienko_PhysRevB.73.094434}, Dzyaloshinskii-Moriya interaction and Kitaev-like anisotropies~\cite{Itamar_PhysRevB.89.014414, Morita_PhysRevB.98.134437}.

\textit{Conclusion---} 
Based on  VMC calculations, we identify four distinct $U(1)$ QSLs: the Dirac spin liquid, Dirac CSL, gapped CSL, and staggered CSL. The staggered CSL has been previously studied in the large $J_{\chi}$ limit, proposing a $U(1)$ gauge flux of $\pm \pi/2$ through up and down triangles~\cite{10.21468/SciPostPhys.13.3.050, PhysRevB.99.035155}. We confirm that this flux pattern minimizes the energy by treating the $U(1)$ gauge flux as a variational parameter. In the intermediate $J_{\chi}$ regime, a new Dirac CSL phase emerges, characterized by time-reversal symmetry-breaking Dirac spinons. Our gapped CSL also exhibits symmetry and flux patterns distinct from those reported in earlier works,~\cite{PhysRevB.91.075112, Gong2014, PhysRevLett.112.137202}, due to the competing $J_{d}$ and $J_{\chi}$ terms. 
Additionally, the interaction between $J_{d}$ and $J_{\chi}$ generates a tricritical point between Dirac CSL and gapped CSL around $(J_{d}/J_{1}, J_{\chi}/J_{1}) \sim (0.3, 0.2)$. The emergence of the tricritical point is briefly discussed using the symmetry-allowed Landau-Ginzburg theory.
As experimental probes, we propose spin-spin correlation, the spin structure factor, the longitudinal thermal conductance exponent, and electromagnetic response to differentiate between these QSLs. Our work suggests new types of $U(1)$ spin liquids grounded in microscopic spin exchange models, offering distinct possibilities for experimental differentiation.
Finally, while our study focused on a specific subset of $U(1)$ chiral spin liquids connected to the Dirac spin liquid flux pattern, numerous  $Z_{2}$ staggered and Kalmeyer-Laughlin-type chiral spin liquids also exist \cite{PhysRevB.93.094437}. Extending our mean-field and VMC analyses to these $Z_2$ states would provide a stringent test of the robustness of our conclusions and potentially uncover new quantum phases with exotic properties.

\begin{acknowledgments}
We thank Sándor Bordács and Yasir Iqbal for the valuable discussions. 
We acknowledge the financial support provided by the Korean National Research Foundation Grant (2021R1A2C109306013), the Hungarian NKFIH Grant No. K142652, and, in part, the National Science Foundation under Grant No. NSF PHY-1748958 to the Kavli Institute for Theoretical Physics (KITP). This material is based upon work supported by the Air Force Office of Scientific Research under award number FA23862514054.
\end{acknowledgments}

\appendix
\section{Construction of a fully symmetrized projected wave function}
\label{sec:irrep}

As discussed in the main text, twisted boundary conditions break the symmetry of the original Hamiltonian, leading to symmetry-breaking in VMC results. To resolve this, we construct a symmetric projected wave function by superposing wave functions related by point group operations.

Let $g$ and $\tilde{g}$ be two gauge choices related by a point group symmetry. The corresponding mean-field Hamiltonians $H^{g}_{\text{MF}}$ and $H^{\tilde{g}}_{\text{MF}}$ are connected via a $U(1)$ gauge transformation $G$ such that $G H^{\tilde{g}}_{\text{MF}} G^\dagger = H^g_{\text{MF}}$. Denoting their eigenvector matrices by $V^g$ and $V^{\tilde{g}}$, we define a unitary matrix $D = V^{g\dagger} G V^{\tilde{g}}$ which satisfies
\begin{equation}
    D E D^\dagger = E,
\end{equation}
where $E = \mathrm{diag}(E_1, \dots, E_N)$ is the energy eigenvalue matrix. Assuming a non-degenerate Fermi energy, $D$ can be block-diagonalized as $D = D_{<} \oplus D_{>}$ for states below and above $E_F$, respectively. The Gutzwiller-projected wave functions $\ket{\Psi^g}$ and $\ket{\Psi^{\tilde{g}}}$ constructed from $V^g$ and $V^{\tilde{g}}$ then satisfy
\begin{equation}
    \ket{\Psi^{\tilde{g}}} = (\det G)^{-1} (\det D_{<})^2 \ket{\Psi^g}.
\end{equation}

To construct a symmetric wave function, we consider four boundary conditions: [PP], [AP], [PA], and [AA], representing periodic (P) or anti-periodic (A) conditions along the primitive vectors $\bm{a}_1$ and $\bm{a}_2$. These boundary conditions are related by the $C_6$ generator of the $C_{6h}$ point group. For example, $C_6$ maps [AP] to [PA] and [AA], while altering the gauge structure. Using the relation above, the $C_6$ operator acts as
\begin{equation}
    C_6
    \begin{pmatrix}
        \ket{\Psi_{\text{PA}}^{g}} \\ 
        \ket{\Psi_{\text{AP}}^{g}} \\ 
        \ket{\Psi_{\text{AA}}^{g}}
    \end{pmatrix}
    =
    \begin{pmatrix}
        0 & e^{i\phi_1} & 0 \\
        0 & 0 & e^{i\phi_2} \\
        e^{i\phi_3} & 0 & 0
    \end{pmatrix}
    \begin{pmatrix}
        \ket{\Psi_{\text{PA}}^{g}} \\ 
        \ket{\Psi_{\text{AP}}^{g}} \\ 
        \ket{\Psi_{\text{AA}}^{g}}
    \end{pmatrix},
\end{equation}
where the phases satisfy $\phi_1 + \phi_2 + \phi_3 = 0 \pmod{2\pi}$. The linear combination
\begin{equation}
    \ket{\Psi_{A_1}^{g}} = \ket{\Psi_{\text{PA}}^{g}} + e^{i\phi_1} \ket{\Psi_{\text{AP}}^{g}} + e^{i(\phi_1 + \phi_2)} \ket{\Psi_{\text{AA}}^{g}}
\end{equation}
is invariant under $C_{6h}$ up to a global phase and thus belongs to the trivial representation. We use this fully symmetric wave function in our VMC calculations.


\section{Magnetoelectric coupling}
\label{sec:magnetoelectric}

\begin{figure}
    \centering
    \includegraphics[width=0.9\columnwidth]{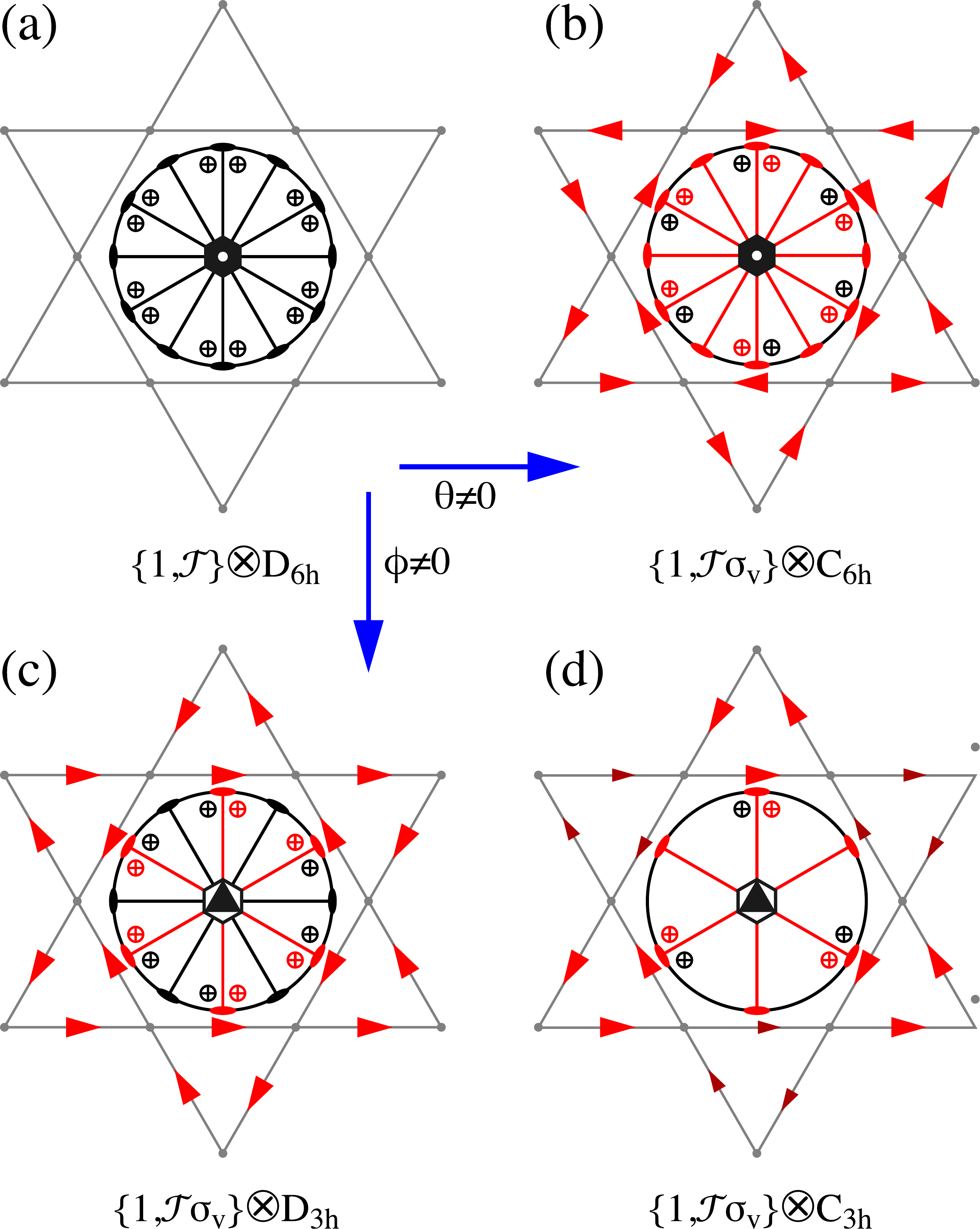}
    \caption{
    Diagrams illustrating the reduction of the grey magnetic point group symmetry $\{1,\mathcal{T}\} \otimes \mathsf{D_{6h}}$ in spin liquids with different flux patterns, indicated by red arrows on the edges of the kagome lattice. Panel (a) shows the full grey point group, while panels (b)–(d) correspond to symmetry reductions induced by finite fluxes: $\theta\neq 0$  in (b) and (d), and a staggered flux ($\phi\neq 0$)  in (c) and (d). The stereograms inside the hexagons represent the magnetic point group associated with each flux configuration. Black symbols denote elements of the unitary subgroup, whereas red symbols indicate elements of the antiunitary subgroup, combining point-group symmetry operations with time-reversal symmetry $\mathcal{T}$.
 }
    \label{fig:magnetic_point_groups}
\end{figure}

Here, we derive the symmetry-allowed interactions between magnetic and electric polarizations in different phases, known as magnetoelectric couplings. To this end, let us extend the two-dimensional point group of the kagome lattice to the three-dimensional group $\mathsf{D_{6h}}$ having 24 elements. A convenient set of generators includes the sixfold rotation $C_6$, the horizontal mirror $\sigma_h\equiv \sigma_{xy}$, and the vertical mirror $\sigma_d \equiv \sigma_{xz}$ (reflection about the $y$ axis). To properly account for magnetization, we also include the antiunitary time-reversal symmetry $\mathcal{T}$.

The Dirac spin liquid preserves all symmetries and is therefore characterized by the grey magnetic point group $\{1,\mathcal{T}\}\otimes\mathsf{D_{6h}}$, see Fig.~\ref{fig:magnetic_point_groups}. 
The Dirac CSL, the staggered CSL, and the Hamiltonian (\ref{eq:J1JdJcHamiltonian}) at finite $J_{\chi}$ are invariant under the black-and-white point group  
$\{1,\mathcal{T}\sigma_v\} \otimes \mathsf{D_{3h}} \equiv \mathsf{D_{6h}}(\mathsf{D_{3h}})$. 
The gapped CSL instead exhibits the symmetry $\{1,\mathcal{T}\sigma_v\}\otimes\mathsf{C_{6h}}\equiv\mathsf{D_{6h}}(\mathsf{C_{6h}})$ for $\phi=0$, which is reduced to $\{1,\mathcal{T}\sigma_v\}\otimes\mathsf{C_{3h}}\equiv\mathsf{D_{3h}}(\mathsf{C_{3h}})$ at finite $\phi$.
All chiral flux patterns considered here are invariant under the combined antiunitary operation $\mathcal{T}\sigma_v$ and the unitary point group $\mathsf{C_{3h}}$.

By considering the transformations of the components of the uniform magnetization $\mathbf{M}$ and electric polarization $\mathbf{P}$, we can construct invariant multinomials for each symmetry group. For example, time reversal $\mathcal{T}$ reverses magnetization, $\mathbf{M} \to -\mathbf{M}$, but leaves polarization unchanged, $\mathbf{P} \to \mathbf{P}$. The combined operation $\mathcal{T}\sigma_v$ transforms $(M_x, M_y, M_z) \to (-M_x, M_y, M_z)$ and $(P_x, P_y, P_z) \to (-P_x, P_y, P_z)$. The reflection $\sigma_h$ transforms $(M_x, M_y, M_z) \to (-M_x, -M_y, M_z)$ and $(P_x, P_y, P_z) \to (P_x, P_y, -P_z)$, and so on. Using these transformations, we systematically construct all symmetry-allowed invariants for each phase and listed them in the last few rows of Table~\ref{tab:tab}.

\section{Fermi lines in the staggered CSL}
\label{sec:fermilines}

Let us briefly discuss how the anti-commutation relation $\{\hat{\mathcal{H}}_{\text{MF}}, \sigma_{v}\} = 0$ in the staggered CSL leads to the emergence of line Fermi surfaces.
We choose the hopping phases of all nearest-neighbors as $\text{arg}(t_{ij}) = \pm \pi/2$ to satisfy $\Phi_{\vartriangle / \triangledown} = \pm\phi = \pm\pi/2$.
Since the reflection symmetry  $\sigma_{v}$ reverses the staggered flux $\phi$, it correspondingly flips the hopping amplitudes, $\sigma_{v}: e^{\pm i\pi/2} = \pm i \to e^{\mp i\pi/2} = \mp i$. 
Therefore, the staggered CSL mean-field Hamiltonian satisfies $\sigma_{v}\hat{\mathcal{H}}_{\text{MF}}\sigma_{v} = -\hat{\mathcal{H}}_{\text{MF}}$, or equivalently $\{\hat{\mathcal{H}}_{\text{MF}}, \sigma_{v}\} = 0$. 
Then, for a $\mathcal{\hat{H}}_{\text{MF}} = \sum_{\bm{k},\sigma}\varepsilon(\bm{k})\hat{f}_{\bm{k}\sigma}^{\dagger}\hat{f}_{\bm{k}\sigma}$, the anti-commutation relation implies 
\begin{align}
    \sigma_{v}\hat{\mathcal{H}}_{\text{MF}}\sigma_{v} = \sum_{\bm{k},\sigma}\varepsilon(\sigma_{v}(\bm{k}))\hat{f}_{\bm{k}\sigma}^{\dagger}\hat{f}_{\bm{k}\sigma} = -\sum_{\bm{k},\sigma}\varepsilon(\bm{k})\hat{f}_{\bm{k}\sigma}^{\dagger}\hat{f}_{\bm{k}\sigma}
\end{align}
for every crystal momentum $\bm{k}$. Therefore, along the vertical mirror plane, where $\sigma_{v}(\bm{k}) = \bm{k}$, we have $\varepsilon(\sigma_{v}(\bm{k})) = \varepsilon(\bm{k}) = -\varepsilon(\bm{k}) = 0$. 
Moreover, the $C_{3}$ rotational symmetry of the staggered CSL ensures the presence of three symmetry-related Fermi lines, all protected by the anti-commutation relation.

\section{Tricritical point}
\label{sec:tricritical}
  
The continuous phase transition between the Dirac chiral spin liquid (CSL) and the gapped CSL becomes discontinuous at the tricritical point, marked by the black circle in the phase diagram of Fig.~\ref{fig:KagomeLattice}(b). The transition is associated with the breaking of the reflection symmetry $\sigma_{d}$ and occurs at a finite value of $\phi$, such that the combined antiunitary symmetry $\mathcal{T}\sigma_v$ remains preserved across the transition. In the following, we analyze this phase transition within a Landau--Ginzburg free-energy framework. 

Imposing the symmetry constraints $\sigma_{d}:(\theta,\phi)\rightarrow (-\theta,\phi)$ and $\mathcal{T}\sigma_v:(\theta,\phi)\rightarrow (\theta,\phi)$, the Landau--Ginzburg free energy involving the uniform ($\theta$) and staggered ($\phi$) fluxes can be written as
\begin{subequations}
    \label{eq:F}
\begin{align}
    \label{eq:Ftot}
    F_{\text{tot}} &= F_{\theta} + F_{\phi} + F_{\theta\&\phi} \\
    \label{eq:Ftheta}
    F_{\theta} &= u_{1}\theta^{6} + f(J_{d},J_{\chi})\theta^{4} - h(J_{d},J_{\chi})\theta^{2}, \\
    \label{eq:Fphi}
    F_{\phi} &= \frac{1}{2}u_{2}\phi^{2} - \alpha J_{\chi}\phi, \\
    \label{eq:Fint}
    F_{\theta\&\phi} &= g(J_{d},J_{\chi})\phi\theta^{2}.
\end{align}
\end{subequations}
Here the coefficients $f(J_{d}, J_{\chi})$, $g(J_{d},J_{\chi})$, and $h(J_{d},J_{\chi})$ are functions of $J_{d}$ and $J_{\chi}$, while $u_{1}$, $u_{2}$, and $\alpha$ are positive constants. The minimal coupling between  $J_{\chi}$ and $\phi$ in  $F_{\phi}$ follows from the fact that both quantities are odd under time-reversal symmetry. By integrating out the $\phi$ field, we study the effect of 
$\phi$ fluctuations on the order of the phase transition through the effective free energy of $\theta$,
\begin{align}
	F_{\theta}^{\text{eff}} &= F_{\theta} - \frac{1}{2u_{2}}\left[\beta J_{\chi} - g\theta^{2}\right]^{2} \equiv u_{1}(\theta^{6} + p\theta^{4} - q\theta^{2}). \nn
\end{align}
The functions $p$ and $q$ are defined as $u_{1}p = f - g^{2}/2u_{2}$ and $u_{1}q = h - \beta J_{\chi}g/u_{2}$. In the $(p,q)$ parameter space, a second-order phase transition occurs when $q$ changes sign from negative to positive with $p > 0$. Conversely, when $p < 0$, the transition becomes first-order, with the boundary given by $q = -p^{2}/4$. Thus, the tricritical point appears at $p=q=0$, indicating that the nature of the phase transition depends on the sign of $p$. This approach effectively describes the VMC phase diagram near the tricritical point.

\bibliography{sample.bib}

@article{PhysRevLett.119.077206,
  title = {Optical Magnetoelectric Resonance in a Polar Magnet $(\mathrm{Fe},\mathrm{Zn}{)}_{2}{\mathrm{Mo}}_{3}{\mathrm{O}}_{8}$ with Axion-Type Coupling},
  author = {Kurumaji, T. and Takahashi, Y. and Fujioka, J. and Masuda, R. and Shishikura, H. and Ishiwata, S. and Tokura, Y.},
  journal = {Phys. Rev. Lett.},
  volume = {119},
  issue = {7},
  pages = {077206},
  numpages = {5},
  year = {2017},
  month = {Aug},
  publisher = {American Physical Society},
  doi = {10.1103/PhysRevLett.119.077206},
  url = {https://link.aps.org/doi/10.1103/PhysRevLett.119.077206}
}

@article{Itamar_PhysRevB.89.014414,
	author = {Kimchi, Itamar and Vishwanath, Ashvin},
	doi = {10.1103/PhysRevB.89.014414},
	issue = {1},
	journal = {Phys. Rev. B},
	month = {Jan},
	numpages = {13},
	pages = {014414},
	publisher = {American Physical Society},
	title = {Kitaev-Heisenberg models for iridates on the triangular, hyperkagome, kagome, fcc, and pyrochlore lattices},
	url = {https://link.aps.org/doi/10.1103/PhysRevB.89.014414},
	volume = {89},
	year = {2014}}

@article{Morita_PhysRevB.98.134437,
	author = {Morita, Katsuhiro and Kishimoto, Masanori and Tohyama, Takami},
	doi = {10.1103/PhysRevB.98.134437},
	issue = {13},
	journal = {Phys. Rev. B},
	month = {Oct},
	numpages = {6},
	pages = {134437},
	publisher = {American Physical Society},
	title = {Ground-state phase diagram of the Kitaev-Heisenberg model on a kagome lattice},
	url = {https://link.aps.org/doi/10.1103/PhysRevB.98.134437},
	volume = {98},
	year = {2018}}

@article{Sergienko_PhysRevB.73.094434,
	author = {Sergienko, I. A. and Dagotto, E.},
	doi = {10.1103/PhysRevB.73.094434},
	issue = {9},
	journal = {Phys. Rev. B},
	month = {Mar},
	numpages = {5},
	pages = {094434},
	publisher = {American Physical Society},
	title = {Role of the Dzyaloshinskii-Moriya interaction in multiferroic perovskites},
	url = {https://link.aps.org/doi/10.1103/PhysRevB.73.094434},
	volume = {73},
	year = {2006}}

@article{Katsura_PhysRevLett.95.057205,
	author = {Katsura, Hosho and Nagaosa, Naoto and Balatsky, Alexander V.},
	doi = {10.1103/PhysRevLett.95.057205},
	issue = {5},
	journal = {Phys. Rev. Lett.},
	month = {Jul},
	numpages = {4},
	pages = {057205},
	publisher = {American Physical Society},
	title = {Spin Current and Magnetoelectric Effect in Noncollinear Magnets},
	url = {https://link.aps.org/doi/10.1103/PhysRevLett.95.057205},
	volume = {95},
	year = {2005}}

@article{Bulaevskii_PhysRevB.78.024402,
	author = {Bulaevskii, L. N. and Batista, C. D. and Mostovoy, M. V. and Khomskii, D. I.},
	doi = {10.1103/PhysRevB.78.024402},
	issue = {2},
	journal = {Phys. Rev. B},
	month = {Jul},
	numpages = {9},
	pages = {024402},
	publisher = {American Physical Society},
	title = {Electronic orbital currents and polarization in Mott insulators},
	url = {https://link.aps.org/doi/10.1103/PhysRevB.78.024402},
	volume = {78},
	year = {2008}}

@article{Murakawa_PhysRevLett.105.137202,
	author = {Murakawa, H. and Onose, Y. and Miyahara, S. and Furukawa, N. and Tokura, Y.},
	doi = {10.1103/PhysRevLett.105.137202},
	issue = {13},
	journal = {Phys. Rev. Lett.},
	month = {Sep},
	numpages = {4},
	pages = {137202},
	publisher = {American Physical Society},
	title = {Ferroelectricity Induced by Spin-Dependent Metal-Ligand Hybridization in ${\mathrm{Ba}}_{2}{\mathrm{CoGe}}_{2}{\mathbf{O}}_{7}$},
	url = {https://link.aps.org/doi/10.1103/PhysRevLett.105.137202},
	volume = {105},
	year = {2010}}

@article{Motrunich_PhysRevB.73.155115,
	author = {Motrunich, Olexei I.},
	doi = {10.1103/PhysRevB.73.155115},
	issue = {15},
	journal = {Phys. Rev. B},
	month = {Apr},
	numpages = {11},
	pages = {155115},
	publisher = {American Physical Society},
	title = {Orbital magnetic field effects in spin liquid with spinon Fermi sea: Possible application to $\ensuremath{\kappa}\text{\ensuremath{-}}{(\mathrm{ET})}_{2}{\mathrm{Cu}}_{2}{(\mathrm{C}\mathrm{N})}_{3}$},
	url = {https://link.aps.org/doi/10.1103/PhysRevB.73.155115},
	volume = {73},
	year = {2006}}

@article{PhysRevB.93.094437,
	author = {Bieri, Samuel and Lhuillier, Claire and Messio, Laura},
	doi = {10.1103/PhysRevB.93.094437},
	issue = {9},
	journal = {Phys. Rev. B},
	month = {Mar},
	numpages = {28},
	pages = {094437},
	publisher = {American Physical Society},
	title = {Projective symmetry group classification of chiral spin liquids},
	url = {https://link.aps.org/doi/10.1103/PhysRevB.93.094437},
	volume = {93},
	year = {2016}}

@article{PhysRevB.91.041124,
	author = {Hu, Wen-Jun and Zhu, Wei and Zhang, Yi and Gong, Shoushu and Becca, Federico and Sheng, D. N.},
	doi = {10.1103/PhysRevB.91.041124},
	issue = {4},
	journal = {Phys. Rev. B},
	month = {Jan},
	numpages = {5},
	pages = {041124},
	publisher = {American Physical Society},
	title = {Variational Monte Carlo study of a chiral spin liquid in the extended Heisenberg model on the kagome lattice},
	url = {https://link.aps.org/doi/10.1103/PhysRevB.91.041124},
	volume = {91},
	year = {2015}}

@article{PhysRevB.92.060407,
	author = {Bieri, Samuel and Messio, Laura and Bernu, Bernard and Lhuillier, Claire},
	doi = {10.1103/PhysRevB.92.060407},
	issue = {6},
	journal = {Phys. Rev. B},
	month = {Aug},
	numpages = {6},
	pages = {060407},
	publisher = {American Physical Society},
	title = {Gapless chiral spin liquid in a kagome Heisenberg model},
	url = {https://link.aps.org/doi/10.1103/PhysRevB.92.060407},
	volume = {92},
	year = {2015}}

@article{PhysRevLett.98.117205,
	author = {Ran, Ying and Hermele, Michael and Lee, Patrick A. and Wen, Xiao-Gang},
	doi = {10.1103/PhysRevLett.98.117205},
	issue = {11},
	journal = {Phys. Rev. Lett.},
	month = {Mar},
	numpages = {4},
	pages = {117205},
	publisher = {American Physical Society},
	title = {Projected-Wave-Function Study of the Spin-$1/2$ Heisenberg Model on the Kagom\'e Lattice},
	url = {https://link.aps.org/doi/10.1103/PhysRevLett.98.117205},
	volume = {98},
	year = {2007}}

@article{10.21468/SciPostPhys.13.3.050,
	author = {Fabrizio Oliviero and Jo{\~a}o Augusto Sobral and Eric C. Andrade and Rodrigo G. Pereira},
	doi = {10.21468/SciPostPhys.13.3.050},
	journal = {SciPost Phys.},
	pages = {050},
	publisher = {SciPost},
	title = {{Noncoplanar magnetic orders and gapless chiral spin liquid on the kagome lattice with staggered scalar spin chirality}},
	url = {https://scipost.org/10.21468/SciPostPhys.13.3.050},
	volume = {13},
	year = {2022}}

@article{PhysRevB.99.035155,
	author = {Bauer, Bela and Keller, Brendan P. and Trebst, Simon and Ludwig, Andreas W. W.},
	doi = {10.1103/PhysRevB.99.035155},
	issue = {3},
	journal = {Phys. Rev. B},
	month = {Jan},
	numpages = {22},
	pages = {035155},
	publisher = {American Physical Society},
	title = {Symmetry-protected non-Fermi liquids, kagome spin liquids, and the chiral Kondo lattice model},
	url = {https://link.aps.org/doi/10.1103/PhysRevB.99.035155},
	volume = {99},
	year = {2019}}

@article{10.21468/SciPostPhys.4.1.004,
	author = {Rodrigo G. Pereira and Samuel Bieri},
	doi = {10.21468/SciPostPhys.4.1.004},
	journal = {SciPost Phys.},
	pages = {004},
	publisher = {SciPost},
	title = {{Gapless chiral spin liquid from coupled chains on the kagome lattice}},
	url = {https://scipost.org/10.21468/SciPostPhys.4.1.004},
	volume = {4},
	year = {2018}}

@misc{bauer2014gapped,
	archiveprefix = {arXiv},
	author = {Bela Bauer and Brendan P. Keller and Michele Dolfi and Simon Trebst and Andreas W. W. Ludwig},
	eprint = {1303.6963},
	primaryclass = {cond-mat.str-el},
	title = {Gapped and gapless spin liquid phases on the Kagome lattice from chiral three-spin interactions},
	year = {2014}}

@article{Savary_2017,
	author = {Lucile Savary and Leon Balents},
	doi = {10.1088/0034-4885/80/1/016502},
	journal = {Reports on Progress in Physics},
	month = {nov},
	number = {1},
	pages = {016502},
	publisher = {IOP Publishing},
	title = {Quantum spin liquids: a review},
	url = {https://dx.doi.org/10.1088/0034-4885/80/1/016502},
	volume = {80},
	year = {2016}}

@article{PhysRevB.83.224413,
	author = {Lu, Yuan-Ming and Ran, Ying and Lee, Patrick A.},
	doi = {10.1103/PhysRevB.83.224413},
	issue = {22},
	journal = {Phys. Rev. B},
	month = {Jun},
	numpages = {11},
	pages = {224413},
	publisher = {American Physical Society},
	title = {${\mathbb{Z}}_{2}$ spin liquids in the $S=\frac{1}{2}$ Heisenberg model on the kagome lattice: A projective symmetry-group study of Schwinger fermion mean-field states},
	url = {https://link.aps.org/doi/10.1103/PhysRevB.83.224413},
	volume = {83},
	year = {2011}}

@article{doi:10.1126/science.1201080,
	author = {Simeng Yan and David A. Huse and Steven R. White},
	doi = {10.1126/science.1201080},
	eprint = {https://www.science.org/doi/pdf/10.1126/science.1201080},
	journal = {Science},
	number = {6034},
	pages = {1173-1176},
	title = {Spin-Liquid Ground State of the <i>S</i> = 1/2 Kagome Heisenberg Antiferromagnet},
	url = {https://www.science.org/doi/abs/10.1126/science.1201080},
	volume = {332},
	year = {2011}}

@article{PhysRevLett.109.067201,
	author = {Depenbrock, Stefan and McCulloch, Ian P. and Schollw\"ock, Ulrich},
	doi = {10.1103/PhysRevLett.109.067201},
	issue = {6},
	journal = {Phys. Rev. Lett.},
	month = {Aug},
	numpages = {6},
	pages = {067201},
	publisher = {American Physical Society},
	title = {Nature of the Spin-Liquid Ground State of the $S=1/2$ Heisenberg Model on the Kagome Lattice},
	url = {https://link.aps.org/doi/10.1103/PhysRevLett.109.067201},
	volume = {109},
	year = {2012}}

@article{PhysRevB.84.020407,
	author = {Iqbal, Yasir and Becca, Federico and Poilblanc, Didier},
	doi = {10.1103/PhysRevB.84.020407},
	issue = {2},
	journal = {Phys. Rev. B},
	month = {Jul},
	numpages = {4},
	pages = {020407},
	publisher = {American Physical Society},
	title = {Projected wave function study of ${\mathbb{Z}}_{2}$ spin liquids on the kagome lattice for the spin-$\frac{1}{2}$ quantum Heisenberg antiferromagnet},
	url = {https://link.aps.org/doi/10.1103/PhysRevB.84.020407},
	volume = {84},
	year = {2011}}

@article{PhysRevB.89.020407,
	author = {Iqbal, Yasir and Poilblanc, Didier and Becca, Federico},
	doi = {10.1103/PhysRevB.89.020407},
	issue = {2},
	journal = {Phys. Rev. B},
	month = {Jan},
	numpages = {5},
	pages = {020407},
	publisher = {American Physical Society},
	title = {Vanishing spin gap in a competing spin-liquid phase in the kagome Heisenberg antiferromagnet},
	url = {https://link.aps.org/doi/10.1103/PhysRevB.89.020407},
	volume = {89},
	year = {2014}}

@article{PhysRevB.87.060405,
	author = {Iqbal, Yasir and Becca, Federico and Sorella, Sandro and Poilblanc, Didier},
	doi = {10.1103/PhysRevB.87.060405},
	issue = {6},
	journal = {Phys. Rev. B},
	month = {Feb},
	numpages = {5},
	pages = {060405},
	publisher = {American Physical Society},
	title = {Gapless spin-liquid phase in the kagome spin-$\frac{1}{2}$ Heisenberg antiferromagnet},
	url = {https://link.aps.org/doi/10.1103/PhysRevB.87.060405},
	volume = {87},
	year = {2013}}

@article{PhysRevB.91.075112,
	author = {Gong, Shou-Shu and Zhu, Wei and Balents, Leon and Sheng, D. N.},
	doi = {10.1103/PhysRevB.91.075112},
	issue = {7},
	journal = {Phys. Rev. B},
	month = {Feb},
	numpages = {9},
	pages = {075112},
	publisher = {American Physical Society},
	title = {Global phase diagram of competing ordered and quantum spin-liquid phases on the kagome lattice},
	url = {https://link.aps.org/doi/10.1103/PhysRevB.91.075112},
	volume = {91},
	year = {2015}}

@article{Gong2014,
	author = {Gong, Shou-Shu and Zhu, Wei and Sheng, D. N.},
	day = {10},
	doi = {10.1038/srep06317},
	issn = {2045-2322},
	journal = {Scientific Reports},
	month = {Sep},
	number = {1},
	pages = {6317},
	title = {Emergent Chiral Spin Liquid: Fractional Quantum Hall Effect in a Kagome Heisenberg Model},
	url = {https://doi.org/10.1038/srep06317},
	volume = {4},
	year = {2014}}

@article{PhysRevLett.112.137202,
	author = {He, Yin-Chen and Sheng, D. N. and Chen, Yan},
	doi = {10.1103/PhysRevLett.112.137202},
	issue = {13},
	journal = {Phys. Rev. Lett.},
	month = {Apr},
	numpages = {5},
	pages = {137202},
	publisher = {American Physical Society},
	title = {Chiral Spin Liquid in a Frustrated Anisotropic Kagome Heisenberg Model},
	url = {https://link.aps.org/doi/10.1103/PhysRevLett.112.137202},
	volume = {112},
	year = {2014}}

@article{PhysRevB.92.125122,
	author = {Wietek, Alexander and Sterdyniak, Antoine and L\"auchli, Andreas M.},
	doi = {10.1103/PhysRevB.92.125122},
	issue = {12},
	journal = {Phys. Rev. B},
	month = {Sep},
	numpages = {6},
	pages = {125122},
	publisher = {American Physical Society},
	title = {Nature of chiral spin liquids on the kagome lattice},
	url = {https://link.aps.org/doi/10.1103/PhysRevB.92.125122},
	volume = {92},
	year = {2015}}

@article{PhysRevB.91.104418,
	author = {Kolley, F. and Depenbrock, S. and McCulloch, I. P. and Schollw\"ock, U. and Alba, V.},
	doi = {10.1103/PhysRevB.91.104418},
	issue = {10},
	journal = {Phys. Rev. B},
	month = {Mar},
	numpages = {8},
	pages = {104418},
	publisher = {American Physical Society},
	title = {Phase diagram of the ${J}_{1}\text{\ensuremath{-}}{J}_{2}$ Heisenberg model on the kagome lattice},
	url = {https://link.aps.org/doi/10.1103/PhysRevB.91.104418},
	volume = {91},
	year = {2015}}

@article{MOUSAVI20163823,
	author = {Hamze Mousavi and Jabbar Khodadadi and Jamshid {Moradi Kurdestany} and Zahra Yarmohammadi},
	doi = {https://doi.org/10.1016/j.physleta.2016.09.043},
	issn = {0375-9601},
	journal = {Physics Letters A},
	number = {45},
	pages = {3823-3827},
	title = {Electrical and thermal conductivities of the graphene, boron nitride and silicon boron honeycomb monolayers},
	url = {https://www.sciencedirect.com/science/article/pii/S0375960116310428},
	volume = {380},
	year = {2016}}

@article{Balents2010,
	author = {Balents, Leon},
	day = {01},
	doi = {10.1038/nature08917},
	issn = {1476-4687},
	journal = {Nature},
	month = {Mar},
	number = {7286},
	pages = {199-208},
	title = {Spin liquids in frustrated magnets},
	url = {https://doi.org/10.1038/nature08917},
	volume = {464},
	year = {2010}}

@article{doi:10.1146/annurev-conmatphys-031218-013401,
	author = {Knolle, J. and Moessner, R.},
	doi = {10.1146/annurev-conmatphys-031218-013401},
	eprint = {https://doi.org/10.1146/annurev-conmatphys-031218-013401},
	journal = {Annual Review of Condensed Matter Physics},
	number = {1},
	pages = {451-472},
	title = {A Field Guide to Spin Liquids},
	url = {https://doi.org/10.1146/annurev-conmatphys-031218-013401},
	volume = {10},
	year = {2019}}

@article{doi:10.1126/science.1163196,
	author = {Patrick A. Lee},
	doi = {10.1126/science.1163196},
	eprint = {https://www.science.org/doi/pdf/10.1126/science.1163196},
	journal = {Science},
	number = {5894},
	pages = {1306-1307},
	title = {An End to the Drought of Quantum Spin Liquids},
	url = {https://www.science.org/doi/abs/10.1126/science.1163196},
	volume = {321},
	year = {2008}}

@article{RevModPhys.89.025003,
	author = {Zhou, Yi and Kanoda, Kazushi and Ng, Tai-Kai},
	doi = {10.1103/RevModPhys.89.025003},
	issue = {2},
	journal = {Rev. Mod. Phys.},
	month = {Apr},
	numpages = {50},
	pages = {025003},
	publisher = {American Physical Society},
	title = {Quantum spin liquid states},
	url = {https://link.aps.org/doi/10.1103/RevModPhys.89.025003},
	volume = {89},
	year = {2017}}

@article{Bauer2014,
	author = {Bauer, B. and Cincio, L. and Keller, B. P. and Dolfi, M. and Vidal, G. and Trebst, S. and Ludwig, A. W. W.},
	day = {10},
	doi = {10.1038/ncomms6137},
	issn = {2041-1723},
	journal = {Nature Communications},
	month = {Oct},
	number = {1},
	pages = {5137},
	title = {Chiral spin liquid and emergent anyons in a Kagome lattice Mott insulator},
	url = {https://doi.org/10.1038/ncomms6137},
	volume = {5},
	year = {2014}}

@article{PhysRevLett.99.097202,
	author = {Schroeter, Darrell F. and Kapit, Eliot and Thomale, Ronny and Greiter, Martin},
	doi = {10.1103/PhysRevLett.99.097202},
	issue = {9},
	journal = {Phys. Rev. Lett.},
	month = {Aug},
	numpages = {4},
	pages = {097202},
	publisher = {American Physical Society},
	title = {Spin Hamiltonian for which the Chiral Spin Liquid is the Exact Ground State},
	url = {https://link.aps.org/doi/10.1103/PhysRevLett.99.097202},
	volume = {99},
	year = {2007}}

@article{PhysRevLett.99.247203,
	author = {Yao, Hong and Kivelson, Steven A.},
	doi = {10.1103/PhysRevLett.99.247203},
	issue = {24},
	journal = {Phys. Rev. Lett.},
	month = {Dec},
	numpages = {4},
	pages = {247203},
	publisher = {American Physical Society},
	title = {Exact Chiral Spin Liquid with Non-Abelian Anyons},
	url = {https://link.aps.org/doi/10.1103/PhysRevLett.99.247203},
	volume = {99},
	year = {2007}}

@article{PhysRevB.80.104406,
	author = {Thomale, Ronny and Kapit, Eliot and Schroeter, Darrell F. and Greiter, Martin},
	doi = {10.1103/PhysRevB.80.104406},
	issue = {10},
	journal = {Phys. Rev. B},
	month = {Sep},
	numpages = {11},
	pages = {104406},
	publisher = {American Physical Society},
	title = {Parent Hamiltonian for the chiral spin liquid},
	url = {https://link.aps.org/doi/10.1103/PhysRevB.80.104406},
	volume = {80},
	year = {2009}}

@article{PhysRevLett.110.067208,
	author = {Cincio, L. and Vidal, G.},
	doi = {10.1103/PhysRevLett.110.067208},
	issue = {6},
	journal = {Phys. Rev. Lett.},
	month = {Feb},
	numpages = {5},
	pages = {067208},
	publisher = {American Physical Society},
	title = {Characterizing Topological Order by Studying the Ground States on an Infinite Cylinder},
	url = {https://link.aps.org/doi/10.1103/PhysRevLett.110.067208},
	volume = {110},
	year = {2013}}

@article{PhysRevLett.115.267209,
	author = {He, Yin-Chen and Bhattacharjee, Subhro and Pollmann, Frank and Moessner, R.},
	doi = {10.1103/PhysRevLett.115.267209},
	issue = {26},
	journal = {Phys. Rev. Lett.},
	month = {Dec},
	numpages = {6},
	pages = {267209},
	publisher = {American Physical Society},
	title = {Kagome Chiral Spin Liquid as a Gauged $U(1)$ Symmetry Protected Topological Phase},
	url = {https://link.aps.org/doi/10.1103/PhysRevLett.115.267209},
	volume = {115},
	year = {2015}}

@article{PhysRevB.95.035141,
	author = {Wietek, Alexander and L\"auchli, Andreas M.},
	doi = {10.1103/PhysRevB.95.035141},
	issue = {3},
	journal = {Phys. Rev. B},
	month = {Jan},
	numpages = {6},
	pages = {035141},
	publisher = {American Physical Society},
	title = {Chiral spin liquid and quantum criticality in extended $S=\frac{1}{2}$ Heisenberg models on the triangular lattice},
	url = {https://link.aps.org/doi/10.1103/PhysRevB.95.035141},
	volume = {95},
	year = {2017}}

@article{PhysRevX.10.021042,
	author = {Szasz, Aaron and Motruk, Johannes and Zaletel, Michael P. and Moore, Joel E.},
	doi = {10.1103/PhysRevX.10.021042},
	issue = {2},
	journal = {Phys. Rev. X},
	month = {May},
	numpages = {16},
	pages = {021042},
	publisher = {American Physical Society},
	title = {Chiral Spin Liquid Phase of the Triangular Lattice Hubbard Model: A Density Matrix Renormalization Group Study},
	url = {https://link.aps.org/doi/10.1103/PhysRevX.10.021042},
	volume = {10},
	year = {2020}}

@article{PhysRevLett.116.137202,
	author = {Hickey, Ciar\'an and Cincio, Lukasz and Papi\ifmmode \acute{c}\else \'{c}\fi{}, Zlatko and Paramekanti, Arun},
	doi = {10.1103/PhysRevLett.116.137202},
	issue = {13},
	journal = {Phys. Rev. Lett.},
	month = {Apr},
	numpages = {6},
	pages = {137202},
	publisher = {American Physical Society},
	title = {Haldane-Hubbard Mott Insulator: From Tetrahedral Spin Crystal to Chiral Spin Liquid},
	url = {https://link.aps.org/doi/10.1103/PhysRevLett.116.137202},
	volume = {116},
	year = {2016}}

@article{PhysRevB.77.224413,
	author = {Hermele, Michael and Ran, Ying and Lee, Patrick A. and Wen, Xiao-Gang},
	doi = {10.1103/PhysRevB.77.224413},
	issue = {22},
	journal = {Phys. Rev. B},
	month = {Jun},
	numpages = {23},
	pages = {224413},
	publisher = {American Physical Society},
	title = {Properties of an algebraic spin liquid on the kagome lattice},
	url = {https://link.aps.org/doi/10.1103/PhysRevB.77.224413},
	volume = {77},
	year = {2008}}

@article{PhysRevB.67.115131,
	author = {Paul, Indranil and Kotliar, Gabriel},
	doi = {10.1103/PhysRevB.67.115131},
	issue = {11},
	journal = {Phys. Rev. B},
	month = {Mar},
	numpages = {8},
	pages = {115131},
	publisher = {American Physical Society},
	title = {Thermal transport for many-body tight-binding models},
	url = {https://link.aps.org/doi/10.1103/PhysRevB.67.115131},
	volume = {67},
	year = {2003}}

@article{PhysRevB.69.165105,
	author = {Joura, A. V. and Demchenko, D. O. and Freericks, J. K.},
	doi = {10.1103/PhysRevB.69.165105},
	issue = {16},
	journal = {Phys. Rev. B},
	month = {Apr},
	numpages = {5},
	pages = {165105},
	publisher = {American Physical Society},
	title = {Thermal transport in the Falicov-Kimball model on a Bethe lattice},
	url = {https://link.aps.org/doi/10.1103/PhysRevB.69.165105},
	volume = {69},
	year = {2004}}

@article{Rong-Yang_chiralKagome_2024,
	author = {Sun, Rong-Yang and Jin, Hui-Ke and Tu, Hong-Hao and Zhou, Yi},
	date = {2024/02/03},
	doi = {10.1038/s41535-024-00627-5},
	id = {Sun2024},
	isbn = {2397-4648},
	journal = {npj Quantum Materials},
	number = {1},
	pages = {16},
	title = {Possible chiral spin liquid state in the S = 1/2 kagome Heisenberg model},
	url = {https://doi.org/10.1038/s41535-024-00627-5},
	volume = {9},
	year = {2024}}

@article{PhysRevLett.59.2095,
	author = {Kalmeyer, V. and Laughlin, R. B.},
	doi = {10.1103/PhysRevLett.59.2095},
	issue = {18},
	journal = {Phys. Rev. Lett.},
	month = {Nov},
	numpages = {0},
	pages = {2095--2098},
	publisher = {American Physical Society},
	title = {Equivalence of the resonating-valence-bond and fractional quantum Hall states},
	url = {https://link.aps.org/doi/10.1103/PhysRevLett.59.2095},
	volume = {59},
	year = {1987}}

\end{document}